\documentclass{emulateapj}
\usepackage{graphics}
\usepackage{psfig}
\usepackage{ulem}
\usepackage{longtable}

\def\sn{{SN~1970G}}
\def\E{{\sl Einstein}}
\def\R{{\sl ROSAT}}
\def\A{{\sl ASCA}}
\def\C{{\sl Chandra}}

\def\X{{\sl XMM-Newton}}
\def\H{{\sl HST}}

\def\gs{\mathrel{\mathchoice {\vcenter{\offinterlineskip\halign{\hfil
$\displaystyle##$\hfil\cr>\cr\sim\cr}}}
{\vcenter{\offinterlineskip\halign{\hfil$\textstyle##$\hfil\cr
>\cr\sim\cr}}}
{\vcenter{\offinterlineskip\halign{\hfil$\scriptstyle##$\hfil\cr
>\cr\sim\cr}}}
{\vcenter{\offinterlineskip\halign{\hfil$\scriptscriptstyle##$\hfil\cr
>\cr\sim\cr}}}}}
\def\ls{\mathrel{\mathchoice {\vcenter{\offinterlineskip\halign{\hfil
$\displaystyle##$\hfil\cr<\cr\sim\cr}}}
{\vcenter{\offinterlineskip\halign{\hfil$\textstyle##$\hfil\cr
<\cr\sim\cr}}}
{\vcenter{\offinterlineskip\halign{\hfil$\scriptstyle##$\hfil\cr
<\cr\sim\cr}}}
{\vcenter{\offinterlineskip\halign{\hfil$\scriptscriptstyle##$\hfil\cr
<\cr\sim\cr}}}}}

\begin{document}

\title{Discovery of X-Ray Emission from Supernova 1970G with \C: \\
Filling the Void between Supernovae and Supernova Remnants}

\author{Stefan~Immler\altaffilmark{1} and K. D. Kuntz\altaffilmark{2}}

\affil{Exploration of the Universe Division, 
X-Ray Astrophysics Laboratory, Code 662, \\
NASA Goddard Space Flight Center, Greenbelt, MD 20771, USA}

\altaffiltext{1}{Universities Space Research Association, Code 662, 
NASA Goddard Space Flight Center, Greenbelt, MD 20771, USA}
\altaffiltext{2}{The Henry A. Rowland Department of Physics and Astronomy,
The Johns Hopkins University, 3400 Charles Street, Baltimore, MD 21218, USA}

\shorttitle{{X-Ray Emission from Supernova 1970G}}
\shortauthors{Immler \& Kuntz}

\begin{abstract}

We report on the discovery of X-ray emission from \sn\ in M101, 35~years 
after its outburst, using deep X-ray imaging with the \C\ X-ray observatory.
The \C\ ACIS spectrum shows that the emission is soft ($\ls 2$~keV) and
characteristic for the reverse shock region.
The X-ray luminosity, $L_{0.3-2}=(1.1\pm0.2)\times10^{37}~{\rm ergs~s}^{-1}$, 
is likely caused by the interaction of the supernova (SN) shock with dense 
circumstellar matter. If the material was deposited
by the stellar wind from the progenitor, a mass-loss rate of 
$\dot{M}=(2.6\pm0.4) \times 10^{-5}~M_{\odot}~{\rm yr}^{-1}~(v_{\rm w}/10~{\rm km~s}^{-1})$ 
is inferred. Utilizing the high-resolution \C\ ACIS data of \sn\ and its 
environment, we reconstruct the X-ray lightcurve from previous \R\ HRI, PSPC, 
and \X\ EPIC observations, and find a best-fit linear rate of decline 
of $L \propto t^{-s}$ with index $s = 1.7\pm0.6$ over a period of 12--35~years
after the outburst.
As the oldest SN detected in X-rays, \sn\ allows, for the first time, direct 
observation of the transition from a SN to its supernova remnant (SNR) phase. 

\end{abstract}


\section{Introduction}
\label{introduction}

Since the launch of the \C\ and \X\ X-ray observatories, the number
of supernovae (SNe) detected in their near aftermath has more than
doubled (see Immler \& Lewin 2003 for a review article)\footnote{A 
complete list of X-ray SNe and references are available at
http://lheawww.gsfc.nasa.gov/users/immler/supernovae\_list.html}.
The high-quality X-ray spectra have confirmed the validity of the 
circumstellar interaction models 
(see Fransson, Lundqvist \& Chevalier 1996 and references therein)
which predict a hard spectral component ($\gs 10$~keV) for the 
forward shock emission during the early epoch ($\ls 100$~days) and a
soft thermal component ($\ls 1$~keV) for the reverse shock emission
after the expanding shell has become optically thin. The soft emission
component dominates the X-ray output of the interaction regions due 
to its higher emission measure and higher electron densities. 
This expected `softening' of the X-ray spectrum has been observationally 
confirmed for a number of young SNe, such as SNe 1978K (Schlegel et~al.\ 2004),
1979C (Immler et~al.\ 2005), 1993J (Zimmermann \& Aschenbach 2003),
1999em and 1998S (Pooley et~al.\ 2002).

Where sufficient data are available, X-ray lightcurves could be established 
over significant time scales (e.g., $\approx 25$~years for SNe 1978K and
1979C; Schlegel et~al.\ 2004, Immler et~al.\ 2005; $\approx 8$~years for
1993J; Immler et~al.\ 2001, Zimmermann \& Aschenbach 2003). 
However, little is known about the transition from a young SN into its 
supernova remnant (SNR) phase as there have been only few detections
of intermediate-age (25--300~yrs) SNe in the radio and none in X-rays.
While the X-ray emission of SNe is dominated 
by the interaction of the shock with the ambient circumstellar matter (CSM), 
likely deposited by the progenitor's stellar wind, the emission from a SNR 
is thought to originate in the shocked interstellar medium (ISM). 
At an age of $\approx 35$~years, \sn\ is the oldest SN detected in X-rays and 
closes this gap which allows one, for the first time, to witness the transition 
from a SN to its SNR phase. 

In this paper we report on the \C\ X-ray observation of \sn,
as well as previous X-ray data from \R\ and \X. In \S \ref{data}
we briefly describe the data and analysis thereof and report on the
X-ray spectrum and multi-mission X-ray lightcurve in \S \ref{results}. 
We discuss the results in the context of the CSM interaction model and 
put \sn\ in the context of the evolution from a SN to a SNR
in \S \ref{discussion}, followed by a summary in \S \ref{summary}.

\section{Data Processing and Analysis}
\label{data}
\begin{figure}[t!]
\unitlength1.0cm
    \begin{picture}(8,8.1) 
\put(0.25,0){ \begin{picture}(9,8)
	\psfig{figure=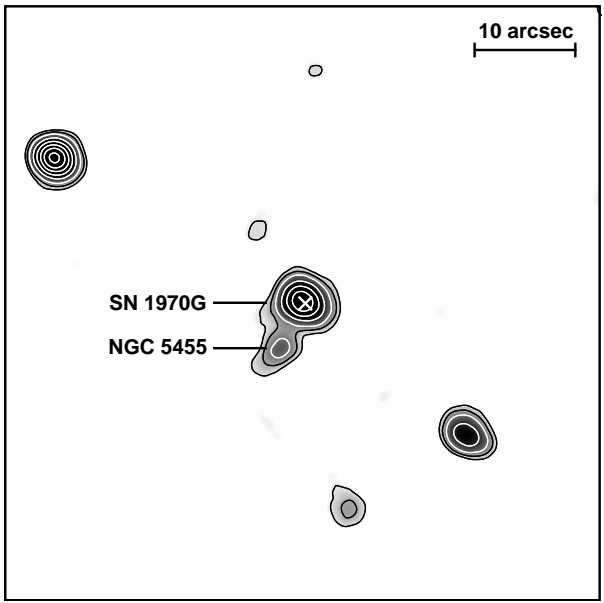,width=8cm,angle=0}
	\end{picture}
	}
    \end{picture}
\caption{
Full (0.3--8~keV) band \C\ ACIS-S2 image of the region (size $1' \times 1'$)
around the position of \sn. The image is smoothed with a Gaussian of
$2\farcs5$ (FWHM) and is given in logarithmic gray-scale ranging from
0 -- 1.5~cts~pix$^{-1}$ (1 pix corresponds to $0\farcs5$). Contour lines are 
0.2, 0.4 (black), 0.6, 0.8, 1, 1.2, 1.4, and 1.6 cts~pix$^{-1}$ (white).
The radio position of \sn\ (Cowan, Goss \& Sramek 1991) north-east of the HII 
region NGC~5455 is marked by a white cross.
} 
\label{fig1}
\end{figure}

\sn\ has been observed with the \C\ observatory as part of the 1~Ms 
observation of the host galaxy M101 (NGC~5457; Kuntz 2005). Five individual 
\C\ ACIS observations (sequence numbers 600389, 600389, 600389, 600389, 600389)
from July 5--11, 2004, were merged into a single observation resulting in 
a cleaned and exposure-corrected on-source exposure time of 139.5~ks for the 
ACIS-S2 chip. Details of the data calibration and analysis are explained in 
Kuntz (2005). The high-resolution ($\approx 0\farcs5$ FWHM) \C\ data were 
used to search for X-ray emission from the position of \sn, and to examine 
the contamination by nearby X-ray sources. A \C\ ACIS-S2 image of \sn\ and 
its environment is given in Fig.~\ref{fig1}.

We further constructed the long-term X-ray lightcurve of 
\sn\ by extracting source counts from previous 
\R\ HRI (May 14 -- Nov 23, 1996; sequence numbers RH600820N00, RH600820A01;
merged exposure time 176.1~ks) and \R\ PSPC observations (Jun 8--9, 1991; 
sequence number RP600108N00; exposure 34.5~ks).
Archival \X\ EPIC observations from 
June 4, 2002 (OBS-ID 0104260101, 43.3~ks)
July 23, 2004 (OBS-ID 0164560701, 41.8~ks)
January 8, 2005 (OBS-ID 0212480201, 32.4~ks)
were also analyzed according to standard analysis procedures described
in the \X\ ABC Guide version 2.01\footnote{available from
http://heasarc.gsfc.nasa.gov/docs/xmm/abc/}. 

A short (2.8~ks) \E\ HRI observation 154 days after the outburst 
of \sn\ did not result in the detection of the SN and is not included in this 
study. Archival \A\ observations were not used due to the large aperture
($\approx 1'$ FWHM) and contamination by nearby X-ray sources.
The \R\ and \X\ count rates were further compared to the published 
results by Wang, Immler \& Pietsch (1999; \R) and Jenkins et~al.\ (2005; \X).

\begin{deluxetable*}{cccccc}
\tabletypesize{\footnotesize}
\tablecaption{X-Ray Properties of SN~1970G \label{tab}}
\tablewidth{0pt}
\tablehead{
\colhead{Epoch} &
\colhead{Instrument} &
\colhead{Count Rate} &
\colhead{$f_{0.3-2}$} &
\colhead{$L_{0.3-2}$} &
\colhead{$\dot{M}$} \\
${\rm [years]}$ & 
& [$10^{-4}~{\rm cts~s}^{-1}$] 
& [$10^{-15}~{\rm ergs~cm}^{-2}~{\rm s}^{-1}$] 
& [$10^{37}~{\rm ergs~s}^{-1}$]
& [$10^{-4}~M_{\odot}~{\rm yr}^{-1}$]} \\
\noalign{\smallskip}
\startdata
11.9 & \R\ PSPC &  $13.0\pm1.3$ & $13.4\pm1.4$ & $8.4\pm0.9$
	& $4.2\pm0.4$ \\
16.8 & \R\ HRI & \phantom{0}$2.9\pm0.6$ & \phantom{0}$6.7\pm1.4$ & $4.2\pm0.9$ 
	& $3.4\pm0.7$ \\
31.9 & \X\ EPIC &  $<48.3$ & $<10.2$ & \phantom{0}$<6.3$ & $<5.8$ \\
34.0 & \X\ EPIC &  $<80.6$ & $<17.0$ & $<10.6$ & $<7.8$ \\
34.5 & \X\ EPIC &  $<119.3$ & $<25.2$ & \phantom{0}$<15.6$ & $<9.4$ \\
34.7 & \C\ ACIS &  $4.3\pm0.6$ & \phantom{0}$1.8\pm0.3$ & $1.1\pm0.2$ 
	& $2.6\pm0.4$
\enddata
\end{deluxetable*}

\section{Results}
\label{results}

A point-like X-ray source is detected at the radio position of \sn\
(Cowan, Goss \& Sramek 1991), with an offset ($\ls 0\farcs5$) similar to the 
spatial resolution and astrometry of the \C\ observation 
(see Fig.~\ref{fig1}). \sn\ is spatially resolved and well separated 
from a nearby ($\approx 5''$) HII region (NGC~5455). 
Exposure, background, and aperture corrected counts extracted from the 
position of \sn\ gives a count rate of 
$(4.3\pm0.6) \times 10^{-4}~{\rm cts~s}^{-1}$.
While the limited photon statistics does not allow a detailed characterization
of the energy distribution of the recorded photons, spectral analysis
shows that the emission is soft, with the bulk of the emission being
confined to the $\ls 2$~keV band. 

Adopting a $kT = 0.6$~keV thermal plasma spectrum typical for the late-time
($\gs100$ days) emission of SNe and assuming a Galactic foreground column 
density with no intrinsic absorption 
($N_{\rm H} = 1.16 \times 10^{20}~{\rm cm}^{-2}$; Dickey \& Lockman 1990) 
gives a 0.3--2~keV flux and luminosity of 
$f_{0.3-2}=(1.8\pm0.3) \times 10^{-15}~{\rm ergs~cm}^{-2}~{\rm s}^{-1}$ and 
$L_{0.3-2}=(1.1\pm0.2) \times 10^{37}~{\rm ergs~s}^{-1}$, respectively, 
for a distance of 7.2~Mpc (Stetson et~al.\ 1998).

Since no information is available about the spectral evolution of \sn\
or other SNe at this age, we used the same spectral template to convert
\R\ and \X\ count rates into fluxes and luminosities to construct the 
long-term (12--35~years) X-ray lightcurve.

An X-ray source is visible in the \R\ HRI images of M101 at the position
of \sn\ (see Fig.~2 in Wang, Immler \& Pietsch 1999). The source, however,
is not included in the \R\ HRI catalog since it is slightly below the employed 
detection threshold (${\rm S/N}=3.5$; Wang, Immler \& Pietsch 1999). Extraction
of source counts within the HRI aperture gives a 0.1--2.4~keV count rate of 
$(2.9\pm0.6)~{\rm cts~s}^{-1}$, corresponding to a 0.3--2~keV band flux and
luminosity of 
$f_{0.3-2}=(6.7\pm1.4) \times 10^{-15}~{\rm ergs~cm}^{-2}~{\rm s}^{-1}$ and 
$L_{0.3-2}=(4.2\pm0.9) \times 10^{37}~{\rm ergs~s}^{-1}$, respectively.

An X-ray source is also listed in the \R\ PSPC catalog at the position of 
\sn\ (source P12; Wang, Immler \& Pietsch 1999), with a signal-to-noise
ratio of ${\rm S/N} = 4.5$. Given that no other X-ray source is visible in the 
high-resolution \C\ image in the environment of \sn\ within the PSPC 
aperture ($\approx 25''$ FWHM), except for low-level emission of a neighboring 
HII region (NGC~5455, see Fig.~\ref{fig1}), it is safe to 
assume that the PSPC source is due to the SN itself. Assuming the emission 
arises from \sn, a (0.5--2~keV band) count rate of 
$(1.3\pm0.1) \times 10^{-3}~{\rm cts~s}^{-1}$ is inferred, corresponding to
$f_{0.3-2}=(1.3\pm0.1) \times 10^{-14}~{\rm ergs~cm}^{-2}~{\rm s}^{-1}$ and 
$L_{0.3-2}=(8.4\pm0.9) \times 10^{37}~{\rm ergs~s}^{-1}$.

A faint X-ray source is visible in each of the nine \X\ EPIC images at low
energies ($<2$~keV). No photons were recorded in the hard (2--6~keV) band.
The source, however, is extended ($\approx 20''$)
and shows no concentration of the recorded photons at the radio position of 
\sn. Since it is likely that the emission is contaminated by the nearby HII 
region and other soft emission components inside the galactic disk, we only 
use the count rates, fluxes, and luminosities as upper limits.
Discrepancies between the published PN and MOS count rates for source 
\#44 (factor $\times 16.5$; Jenkins et~al.\ 2005) for one of the published
\X\ data (OBS-ID 0104260101) makes a comparison problematic.

We calculated the mass-loss rate of the progenitor as a function 
of the stellar wind age using the relationship
$L_{\rm x} = 4/(\pi m^2) \Lambda(T) \times (\dot{M}/v_{\rm w})^2\times(v_{\rm s} t)^{-1}$, where $m$ is the mean mass per particle ($2.1\times10^{-27}$~kg 
for a H+He plasma), $\Lambda(T)$ the cooling function of the heated plasma 
at temperature $T$, $\dot{M}$ the mass-loss rate of the progenitor, 
$v_{\rm w}$ the speed of the stellar wind blown off by the progenitor, 
and $v_{\rm s}$ the speed of the outgoing shock (see Immler, Wilson \&
Terashima 2002).
A constant shock velocity of $v_{\rm s}=9,000~{\rm km~s}^{-1}$ similar 
to other core-collapse SNe, such as SN~1979C (Marcaide et~al.\ 2002) was 
assumed. An effective (0.3--2~keV band) cooling function of 
$\Lambda = 3 \times 10^{-23}~{\rm ergs~cm}^3~{\rm s}^{-1}$ for 
an optically thin thermal plasma with a temperature of $10^7$~K was adopted, 
which corresponds to the assumed thermal plasma temperature.
Assuming different plasma temperatures in the range 0.5--1~keV
would lead to changes in the emission measure of $\ls 10\%$.
Key observational properties of \sn\ are listed in Table~\ref{tab}.

\section{Discussion}
\label{discussion}

Previous X-ray observations of \sn\ with \R, \A, and \X\ lacked the
spatial resolution needed to separate the SN from the nearby ($\approx 5''$)
HII region NGC~5455. Using the sub-arcsec \C\ imaging capability, in
combination with the deep exposure time (140~ks), we detect soft 
($\ls 2$~keV) X-ray emission from \sn\ with a luminosity of 
$L_{0.3-2} \approx 1 \times 10^{37}~{\rm ergs~s}^{-1}$, 35~years after its 
outburst. The source is spatially resolved from the nearby HII region NGC~5455
($\approx 5''$), which is observed at a $\ls20\%$ flux level compared to \sn.
Contamination of the previous \R\ data of \sn\ with the HII region
should therefore be minimal, especially when kept in mind that at earlier 
epochs \sn\ was significantly brighter, while the HII region presumably has 
a constant flux.

Given the information from the high-resolution spatial imaging with \C, 
we recover X-rays from \sn\ in the \R\ HRI and PSPC observations.
\sn\ is not conclusively detected in three recent (2002--2005) \X\ observations
due to the extent of the source ($\approx 20''$) as imaged by the PN and MOS 
instruments, and positional offsets larger than the instrumental uncertainties
($\approx 5''$).  

In combination, an X-ray rate of decline of $L \propto t^{-s}$ with 
index $s = 1.7\pm0.6$ is similar to other X-ray SNe (e.g., SN~1999em, 
Pooley et~al.\ 2002; SN~1986J, Houck et~al.\ 1998). This strengthens the case 
that the emission is due to the SN itself, even though a variable source
in the environment of NGC~5455 cannot be entirely ruled out.
The lack of X-rays above 2~keV confirms that the emission is soft and of 
possible thermal original. Such a soft emission is expected for the late 
emission of a SN, where the shocked emission originates from the reverse 
shock region. 

Assuming that the emission arises from shock heated plasma deposited by
the progenitor's stellar wind, a mass-loss rate of 
$\dot{M}=(2.6\pm0.4) \times 10^{-5}~M_{\odot}~{\rm yr}^{-1}~(v_{\rm w}/10~{\rm km~s}^{-1})$ 
is inferred for the late epoch observed with \C. 
A similar mass-loss rate was inferred from radio observations 
[$2 \times 10^{-5}~M_{\odot}~{\rm yr}^{-1}~(v_{\rm w}/10~{\rm km~s}^{-1})$;
Weiler 1993]. The mass-loss rates for the earlier epoch are slightly
higher and might indicate an evolution in the mass-loss rate history of the 
progenitor (see Tab.~\ref{tab}). It should be noted, however, 
that the lack of knowledge about the shock front velocity precludes a more
detailed study. If the shock front experienced a significant deceleration,
as expected when the shock has swept up a significant amount of CSM,
the lower inferred mass-loss rates for the later epochs would be even lower
(see the discussion about the mass-loss rate history of SN~1979C in 
Immler et~al.\ 2005). However, given the uncertainties in the derived
X-ray luminosities due to a possible residual contamination of the \R\
and \X\ data caused by their larger point-spread functions 
(PSF; $\approx 5''$ FWHM for the \R\ HRI; $5''$--$6''$ FWHM for the \X\ 
EPIC instrument; $\approx 25''$ FWHM for the \R\ PSPC),
the apparent increase of the mass-loss rate at earlier epochs is not regarded
as significant. Interestingly, the mass-loss rates are higher
for the instruments with the largest PSF (EPIC and PSPC).
If the mass-loss rates inferred from the instruments with the smallest PSF
are compared [$\dot{M} = (3.4\pm0.7)$ and 
$(2.6\pm0.4) \times 10^{-5}~M_{\odot}~{\rm yr}^{-1}~(v_{\rm w}/10~{\rm km~s}^{-1})$
for the HRI and ACIS, respectively], no statistically significant differences
are observed. We therefore conclude that \sn\ shows a mass-loss rate which
is consistent with being constant over a period of 12--35~years 
after the peak optical brightness (July 30, 1970; Detre 1970),
corresponding to $\approx$ 11,000--31,000 years in the 
stellar wind age using the conversion $t_{\rm w} = t v_{\rm s}/v_{\rm w}$
and assuming a stellar wind speed of $10~{\rm km~s}^{-1}$ and a shock front
velocity of 9,000$~{\rm km~s}^{-1}$.

The measured mass-loss rate for \sn\ is similar to those inferred for other
Type II SNe, which typically range from $10^{-5}$ 
to $10^{-4}~M_{\odot}~{\rm yr}^{-1}$ (see Immler \& Lewin 2003).
This is indicative that the X-ray emission arises from shock-heated CSM 
deposited by the progenitor rather than shock-heated ISM, even at this late 
epoch after the outburst.

\begin{figure}[t!]
\unitlength1.0cm
    \begin{picture}(3,6.4) 
\put(-0.9,0){ \begin{picture}(1,1)
	\psfig{figure=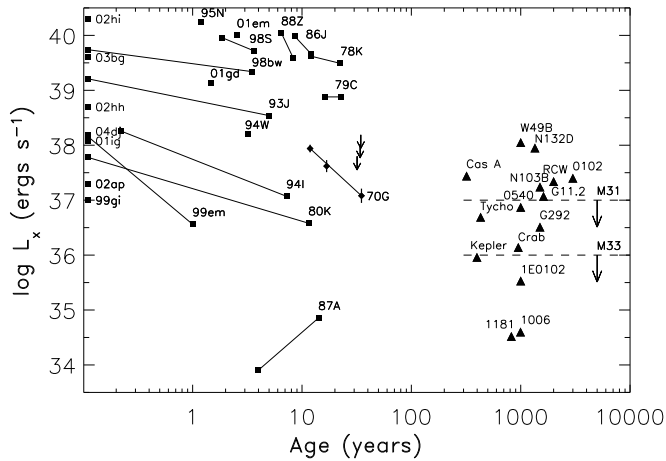,width=9.6cm,angle=0}
	\end{picture}
	}
    \end{picture}
\caption{
Soft (0.3--2~keV) band X-ray luminosities of all SNe detected to date 
(filled squares) and historical SNRs (filled triangles) as a function 
of age (in units of years).
The X-ray lightcurve of \sn\ is marked by filled diamonds with error bars
(from left to right: \R\ PSPC, \R\ HRI and \C\ ACIS). \X\ 
EPIC upper limits are indicated by arrows.
}
\label{fig2}
\end{figure}

In order to explain the X-ray luminosities of SNRs such as Cas A 
($L_{0.3-2}=2.8 \times 10^{37}~{\rm ergs~s}^{-1}$ at an age of
$\approx 320$~years),
mass-loss rates in excess of $10^{-4}~M_{\odot}~{\rm yr}^{-1}$ would be
required in the shocked stellar wind scenario. The high required amounts
of deposited material could not be provided by a sustained stellar wind
since the wind would interact with the ambient ISM at these large radii
($\gs 10^{18}$~cm) from the site of the explosion.
Instead, the X-ray emission from SNRs is thought to originate in radiative
cooling of the shock-heated ISM which has substantially larger densities. 
The relatively low inferred mass-loss rate required to produce the
observed X-ray luminosity of \sn\ indicates that the SN shock has not yet
reached the ISM. Consequently, the \sn\ has not yet reached its SNR phase.

We therefore conclude that at an age of 35~years, \sn\ is the oldest SN 
observed in X-rays, and provides important constraint as to when the 
transition from a SN to its SNR phase occurs, which is likely at a later 
stage ($\approx 50$--100~years) after its outburst.

In Fig.~\ref{fig2} we plot the X-ray evolution of all SNe detected to 
date\footnote{http://lheawww.gsfc.nasa.gov/users/immler/supernovae\_list.html} 
as well as the X-ray luminosities of historical Galactic, LMC, SMC 
(as observed with 
\C\footnote{http://www.astro.psu.edu/users/green/Main/main5.html}), 
M31 (Supper et~al.\ 2001; Kong et~al.\ 2003), and M33 (Ghavamian et~al.\ 2005)
SNRs. A comparison of the X-ray luminosities shows that the transition from
a SN to a SNR appears to be rather smooth. 
If the evolution of \sn\ continues as observed to date, it will be similar 
to X-ray faint Galactic SNRs such
as SNRs 1181 and 1006. X-ray bright SNe such as SNe 1978K and 1979C,
which show no sign for an X-ray evolution out to large distances
($\approx 10^{17}$~cm) at an age of $\approx 25$~years after their
outburst, can evolve into luminous SNRs such as Cas A due to their
strong interaction with their dense ambient CSM, or possibly attain
the X-ray luminosities of the brightest SNRs observed (such as W49B
or N132D) in case of a dense ISM interaction. 

Clearly, X-ray detections and monitoring of intermediate-age (around 100~years) 
SNe are needed to address these questions and to study the transition from a SN
into a SNR in detail. Deep galaxy surveys with \C\ and \X\
and future X-ray missions such as {\sl Const-X} and {\sl XEUS}
provide the capabilities to detect these intermediate-age SNe in the near 
future and to fill the void in the evolution of a SN to a SNR.

\section{Summary}
\label{summary}

Based on our deep (140~ks) X-ray imaging of M101 with the \C\ observatory
we present the detection of \sn\ in X-rays. Key results are:
\\ \\
\noindent 
$\bullet$~\sn\ is detected with a luminosity of 
$L_{0.3-2}=(1.1\pm0.2)\times10^{38}~{\rm ergs~s}^{-1}$,
35 years after its outburst in our deep \C\ observation of M101.

\noindent 
$\bullet$~\sn\ is recovered in previous \R\ HRI and PSPC observations 
and shows a best-fit linear rate of decline of $L \propto t^{-s}$ 
with index $s = 1.7\pm0.6$ over the observed period of $12$--$35$~yrs 
after its outburst, consistent with recent \X\ observations.

\noindent 
$\bullet$~The emission of \sn\ is soft ($\ls 2$~keV) and indicates that
it originates in the reverse shocked region, as expected for the late
emission in the evolution of a SN.

\noindent 
$\bullet$~A mass-loss rate of
$\dot{M}=(2.6\pm0.4) \times 10^{-5}~M_{\odot}~{\rm yr}^{-1}$ 
is inferred consistent with being constant over a period of 
11,000--31,000~years in the evolution of the progenitor.

\noindent 
$\bullet$~At an age of 35 years, the SN has not yet reached its SNR 
phase, which is likely to occur at a later stage in the evolution
($\approx$~50--100~years).

\noindent 
$\bullet$~Comparison of the X-ray lightcurves of all SNe detected to
date with the X-ray luminosities of SNRs indicates a smooth transition 
of SNe into their SNR phases.

\acknowledgments
This paper is based on observations obtained with XMM-Newton, an ESA 
science mission with instruments and contributions directly funded by 
ESA Member States and NASA. 
KDK was supported by the \C\ grant SAO GO-5600587.
The authors wish to thank R. Petre for helpful suggestions and comments.

\end{document}